\documentclass[final]{pasj00}
\usepackage{graphicx}
\SetRunningHead{T. Akahori and K. Yoshikawa}{Non-Equilibrium
  Ionization State and Two-Temperature Structure in the Bullet Cluster}

\title{Non-Equilibrium Ionization State and Two-Temperature Structure
in the Bullet Cluster 1E0657-56}
\author{Takuya \textsc{Akahori}}
\affil{Research Institute of Basic Science, Chungnam National University,
Daejeon 305-764, Korea}
\email{akataku@canopus.cnu.ac.kr}
\author{and \\ Kohji \textsc{Yoshikawa}}
\affil{Center for Computational Sciences, University of Tsukuba, 1-1-1, 
Tennodai, Tsukuba, Ibaraki 305-8577}
\email{kohji@ccs.tsukuba.ac.jp}


\Received{2011 July 15}
\Accepted{2011 September 3}
\Published{2012 Febrary 25 (Vol. 64, No. 1)}

\begin{document}
\maketitle

\begin{abstract}
 We investigate a non-equilibrium ionization state and an electron-ion
 two-temperature structure of the intracluster medium in the merging
 galaxy cluster, 1E0657-56 (the Bullet cluster), using a series of
 N-body and hydrodynamic simulations. We find that the electron 
temperature at the shock layer associated with the X-ray sub peak (bullet)
is quite different depending on the thermal relaxation model between 
electrons and ions; $\sim 25$~keV for the Coulomb thermal relaxation 
model and $\sim 45$~keV for the instantaneous thermal relaxation
 model in the simulations which reproduce the observed X-ray morphology.
Furthermore, both of Fe\,\textsc{xxv} and Fe\,\textsc{xxvi} are
 overabundant compared with the ionization equilibrium state around the
 shock layer, and thus, the intensity ratio between Fe\,\textsc{xxv} and
 Fe\,\textsc{xxvi} K$\alpha$ lines are significantly altered from that
 in the ionization equilibrium state. We also carry out the simulations
 with various sets of merger parameters, and discuss a possible range of
 the non-equilibrium effects in this system. Our results could be tested
 with future X-ray observations such as {\it Astro-H} with better
 sensitivity in high energy band.
\end{abstract}

\section{Introduction}

Cosmological shock heating induced by successive merging of galaxies,
galaxy groups, and clusters, are believed to be the main mechanism
that have heated the X-ray emitting intracluster medium (ICM). 
Hydrodynamic simulations of the structure formation have shown that 
shock waves in galaxy clusters has a Mach number of a few in the 
standard $\Lambda$CDM cosmology (e.g., \cite{ryu03}), 
which heat the ICM at most several tens keV. Such very hot ICM has been 
found through X-ray emission and the Sunyaev-Zel'dovich (SZ) effect 
(e.g., RXC J1347.5-1144: \cite{kit04,ota08}, Abell 3667: \cite{nak09}),
suggesting that violent mergers are ongoing in the context of a standard
scenario of the hierarchical structure formation in the universe.

Cosmological shock waves are also important as laboratories
of collisionless plasma. For instance, turbulent-flow motions induced
via cascade of the vorticity generated at the shocks amplify magnetic
fields through the turbulent dynamo (e.g., \cite{rkcd08}). Related to
the magnetic field, particle acceleration and nonthermal emission are
also interesting phenomena around the shocks (see \cite{sar02} for a
review). Distribution functions for shock heated electrons were recently
studied in the framework of the relativistic correction of the SZ effect
(\cite{pro11a,pro11b,pro11c}). In last decade, a possible difference in
temperature between electrons and ions, or the two-temperature
structure, and the non-equilibrium ionization state of the ICM at the
post-shock regions have been intensively investigated
(\cite{tak99,tak00,yfh05,ys06,cf06,aka08,rn09,aka10,won11}).

\citet{aka10} (hereafter AY10) simulated merging galaxy clusters,
relaxing the assumptions of both an ionization equilibrium and a thermal 
equipartition between electrons and ions, and found that both assumptions
are not justified around the shocks in merging galaxy clusters, because 
their relaxation timescales, order of $10^7$~yr for the electron density of
$10^{-3}$~${\rm cm^{-3}}$, are not short enough compared with the 
merger timescale. They demonstrated that deviations from the ionization
equilibrium and electron-ion equipartition states significantly affect an
interpretation of the observational data such as the ICM temperature and
metallicity. Therefore, studying the non-equilibrium effects is necessary
not only to understand the relaxation mechanism in the plasma but also to
correctly measure the properties of the shock-heated ICM with X-ray
continuum emission of free electrons and X-ray line emissions of heavy
elements. The effects of the non-equilibrium states in galaxy clusters have
already been recognized with {\it Suzaku} X-ray observatory
(\cite{ota08,hos10,aka11}).

1E0657-56 or RXC J0658.5-5556 ($z=0.296$) is known as one of the hottest
galaxy clusters. It hosts an X-ray sub peak on $\sim 500$~kpc west side
from a main peak in the {\it ROSAT} HRI image (\cite{tuk98}). {\it Chandra}
observations (\cite{mar02,mar06}) have revealed that the X-ray sub peak
has a ``bullet''-like shape, that is why 1E0657-56 is called the Bullet
cluster. Except for a region around the sub peak, average ICM
temperature was estimated to be $kT = 17.4 \pm 2.5$~keV (\cite{tuk98})
and $14.1 \pm 0.2$~keV (\cite{mar02}) from the fitting of the {\it ASCA}
and {\it Chandra} data, respectively, which is relatively high compared
with other observed clusters (e.g., \cite{ota06,cav09}). Thermal
SZ effect through such hot ICM was also observed (e.g., \cite{zem10}).
Optical-band observations have shown that galaxy components are clearly 
offset from the associated X-ray peaks (e.g., \cite{bar02}), and the 
gravitational potential does not trace the distribution of the ICM but follows 
approximately the distribution of galaxies as expected for a collisionless 
dark matter component (\cite{clo04,clo06,bra06}). These features suggest 
that a violent merger of a sub cluster is ongoing in the 1E0657-56 system.
A density jump at a bow-shock structure ahead of the bullet indicates a
shock front with a Mach number of $M=3.0\pm 0.4$ (\cite{mar06}). 
Therefore, the 1E0657-56 system is an suitable target to investigate 
the non-equilibrium ionization state and the electron-ion two-temperature
structure. Many theoretical works were devoted to numerical simulations
of the 1E0657-56 system (\cite{tak06,mil07, spr07,nus08,mas08}), and
concluded that an ongoing merger of two galaxy clusters can reproduce
the observed features such as very hot ICM, shocks, and displacement
between X-ray and column density peaks. These works, however, 
have assumed the ionization equilibrium of the ICM and the thermal
equipartition between electrons and ions, and we are the first to
address the non-equilibrium states of the ICM for further elucidation
of the physical properties of the ICM in 1E0657-56 system.

In this paper, we investigate an ionization state and temperature
structure of the ICM in the 1E0657-56 system, and clarify to what extent
deviations from ionization equilibrium and thermal equipartition
between electrons and ions arise and affect an interpretation of the
observational data. We also discuss future prospects for the
detectability of these non-equilibrium phenomena using the near future
X-ray missions such as {\it Astro-H}.

The rest of this paper is organized as follows. In section 2, we describe
the model and setup of the simulations. The results are shown in section 3.
We discuss parameter dependence of our results and effect of radiative cooling
in section 4 and 5, respectively, and summarize our conclusions in section 6. 
In the case that the cosmological scaling is required, we assume the density
parameter, $\Omega_{\rm M}=0.24$, the cosmological constant, 
$\Omega_{\Lambda}=0.76$, the baryon density parameter, 
$\Omega_{\rm b}=0.04$, and the Hubble constant,
$H_0 = 70$ km s$^{-1}$ Mpc$^{-1}$, unless otherwise specified.

\section{Numerical Simulation}

We carry out a set of N-body and SPH simulations of the 1E0657-56
system, relaxing the assumptions of the ionization equilibrium of
the ICM and the thermal equipartition between electrons and ions.
Detailed descriptions of our simulation code were already published in AY10.
Below, we briefly summarize our numerical schemes to treat the non-equilibrium
ionization state and the thermal relaxation between electrons and ions
as well as the setup of our simulations.

\subsection{Electron-Ion Two Temperature Structure}

Timescales on which each of electrons and ions achieves thermal
relaxation through Coulomb scattering are much shorter than that between
electrons and ions and a timescale of the merger in the cluster
environment. Thus, we assume that electrons and ions always reach
Maxwellian distributions with temperatures, $T_{\rm e}$ and $T_{\rm i}$,
respectively. Here, electrons and ions can have different temperatures
just after experiencing shock heating, since most of the kinetic energy
of the ICM is carried by ions at a shock front, and is preferentially
converted to the thermal energy of ions in the post-shock regions
(\cite{fox97}). Afterward, thermal relaxation between electrons and ions
proceeds toward the equipartition state. 

Physical processes responsible for the thermal relaxation are still
matter of debate. It is well known that the Coulomb scattering is not
efficient enough to achieve electron--ion thermal equipartition in a
timescale sufficiently shorter than typical merger timescale, $\simeq
10^8~{\rm yr}$, and it is argued that plasma waves are able to attain
the thermal relaxation between electrons and ions more quickly than the
Coulomb scattering. In this work, in order to bracket the plausible
range of theoretical uncertainties regarding the thermal relaxation
processes between electrons and ions, we present the results with two
different thermal relaxation models: ``single-temperature runs'' in which the
equipartition of thermal energy between electrons and ions is achieve
instantaneously, and ``two-temperature runs'' which adopt only the
Coulomb scattering as a physical process for thermal relaxation between
electrons and ions.

In the two-temperature runs, we solve an ordinary energy equation for
mixed fluid with a mean temperature, $\bar{T}\equiv (n_{\rm e}T_{\rm
e}+n_{\rm i}T_{\rm i})/(n_{\rm e}+n_{\rm i})$, where $n_{\rm e}$ and
$n_{\rm i}$ the number density of electrons and ions, respectively, and
another equation for the time evolution of electrons which can be
reduced to 
\begin{equation}\label{eq:2T}
\frac{d\tilde{T}_{\rm e}}{dt}=
\frac{\tilde{T}_{\rm i}-\tilde{T}_{\rm e}}{t_{\rm ei}}
-\frac{\tilde{T}_{\rm e}}{u}Q_{\rm sh},
\end{equation}
where $\tilde{T}_{\rm e}\equiv T_{\rm e}/\bar{T}$ and $\tilde{T}_{\rm
i}\equiv T_{\rm i}/\bar{T}$ are the dimensionless temperatures of
electrons and ions, respectively, normalized by the mean temperature,
$t_{\rm ei}$ the Coulomb thermal relaxation timescale between electrons
and ions (equation (3) of AY10), and $u$ and $Q_{\rm sh}$ are the
specific thermal energy and the shock heating rate per unit mass,
respectively. We solve equation~(\ref{eq:2T}) in the same manner
described in \citet{tak99}.

The effect of radiative cooling of the ICM is considered in our
simulations, in which only the thermal bremsstrahlung emission is
considered since the ICM temperature is always high enough ($\gg$ a
few~keV). We compute the cooling rate by using $\bar{T}$ in the
single-temperature runs and $T_{\rm e}$ in the two-temperature runs to
incorporate the effect of two-temperature structure of the ICM
consistently.

\subsection{Non-Equilibrium Ionization State}

The Lagrangian time evolution of ionization fractions of heavy elements, 
C, N, O, Ne, Mg, Si, S, and Fe, as well as hydrogen and helium is 
computed for each SPH particle by
solving rate equations:
\begin{equation}\label{eq:NEI}
\frac{df_j}{dt}=\sum_{k=1}^{j-1}S_{j-k,k}f_k-
\sum_{i=j+1}^{Z+1}S_{i-j,j}f_j-\alpha_jf_j+\alpha_{j+1}f_{j+1},
\end{equation}
where $j$ is the index of a particular ionization stage considered, $Z$
the atomic number, $f_j$ the ionization fraction of an ion $j$,
$S_{i,j}$ the ionization rate of an ion $j$ with the ejection of $i$
electrons, and $\alpha_j$ is the recombination rate of an ion $j$. The
ionization processes include collisional, Auger, charge-transfer, and
photo-ionizations, and recombination processes are composed of radiative
and dielectronic recombinations. We solve equation~(\ref{eq:NEI}) in the
same manner described in \citet{ys06}.

Ionization and recombination rates are calculated by utilizing the SPEX
ver 1.10 software package\footnote{ $\langle$
http://www.sron.nl/divisions/hea/spex/ $\rangle$.}. In the
two-temperature runs, the reaction rates are computed using the electron
temperature, $T_{\rm e}$, calculated from equation~(\ref{eq:2T}) in
order to incorporate the effect of the two-temperature structure. As for
the single-temperature runs, the reaction rates are computed using the
mean temperature, $\bar{T}$. All ionization states of the atoms denoted
above are considered in calculations of X-ray surface brightness. But in
the following we only focus on the ionization state of iron, because
most of ions except iron are fully-ionized for the ICM with the
temperature higher than several keV in the 1E0657-56 system.

\subsection{Simulation Setup}

As an initial condition of the 1E0657--56 system, two spherically
symmetric galaxy clusters are set up as follows.  The virial radius,
$r_{\rm 200}$, within which the mean mass density is 200 times the
present cosmic critical density, and the virial mass, $M_{200}$,
enclosed within $r_{\rm 200}$ of a main (massive) cluster are set to
$r_{\rm 200}=2.36$~Mpc and $M_{\rm 200} = 1.5 \times 10^{15}$~$\MO$,
respectively, and those of a sub (less massive) cluster are $1.21$~Mpc
and $2.5 \times10^{14}$~$\MO$, respectively. We adopt the hydrostatic
model (AY10) with the NFW density profile \citep{nfw97} for a dark
matter halo, and the $\beta$-model density profile \citep{cav76} with
$\beta = 2/3$ for an ICM component, and both of dark matter and SPH
particles are initially distributed out to $r_{\rm 200}$.  The scale
radii, $r_{\rm s}$, of the NFW profile are set so that the concentration
parameters $c=r_{\rm s}/r_{\rm 200}=$ 5 and 8 for the main and sub
clusters, respectively, and the core radius of the $\beta$-model is set
to $r_{\rm c}=r_{\rm s}/2$ for each cluster.

The main cluster is composed of $10^7$ dark matter particles and the
same number of SPH particles, while those of the sub clusters are
proportional to their mass ratio. The corresponding spatial resolution
of SPH is 12~kpc at the ICM density of $10^{-2}$~${\rm cm^{-3}}$ (at the
bullet) and 26~kpc at $10^{-3}$~${\rm cm^{-3}}$ (at the shock front).

Our nominal run that reproduces the gross observed features of the
1E0657-56 system is organized as follows. Initially, two galaxy clusters
contact each other at their outer edge, $r_{\rm 200}$. Their initial
relative velocity is set to 3000~km/s, and the impact parameter is
$b=0.236$~Mpc. Such an initial relative velocity and a non-zero impact
parameter are required to reproduce a shock front with a Mach number of
$M\sim 3$ and asymmetric X-ray surface brightness (section 4; see also 
\cite{mas08}). As for the metallicity, a spatially uniform metallicity of 0.2 
times solar abundance is assumed. In addition to that, we also assume
$\tilde{T}_{\rm e}=\tilde{T}_{\rm i}=1$ and an ionization equilibrium
state at the start of the simulations.

Compared with the previous hydrodynamic simulations \citep{spr07,mas08},
the numbers of SPH and dark matter particles are $\sim$5--10 times
larger, so that the spatial resolution is roughly twice better. 
While \citet{mas08} adopted the larger concentration parameters
and smaller main cluster mass, we adopt the larger concentration parameters
but the same main cluster mass as those adopted in \citet{spr07}. This is 
because our parameters better reproduce the observed temperature profile
at the pre-shock region in front of the bullet.
While \citet{spr07} neglected the effect of radiative cooling, we demonstrate
that considering radiative cooling is important since the radiative cooling
significantly affects the width between the shock front and the contact
discontinuity associated with ``the bullet'' (section 5). Although our 
nominal run reproduces an overall structure of the 1E0657-56 system well, 
as a practical interest, we also carried out the simulations with other sets 
of parameters, and discuss a possible range of the non-equilibrium effects 
(section 4). 

\section{Result}

In the rest of this paper, we present the result after 1.12~Gyr has
elapsed from the initial condition. At this epoch, the sub cluster
already penetrates the center of the main cluster, and the separation
between the centers of the main and sub clusters reaches $\sim$
0.72~Mpc, which is consistent with that of the 1E0657-56 system (e.g.,
\cite{clo04}), assuming that the collision plane is perpendicular to the
line-of-sight (the line-of-sight velocity difference of galaxies is
estimated to be only $600$~km/s; \cite{bar02}).

\subsection{Overall Structure of Simulated 1E0657-56}

We first present the overall structure of the simulated clusters in
order to make sure that our simulation reproduces the gross observed
features of the 1E0657-56 system, and to locate the regions where the
effects of the non-equilibrium ionization state and the two-temperature
structure remarkably take place.

\begin{figure}[tp]
\begin{center}
\FigureFile(80mm,40mm){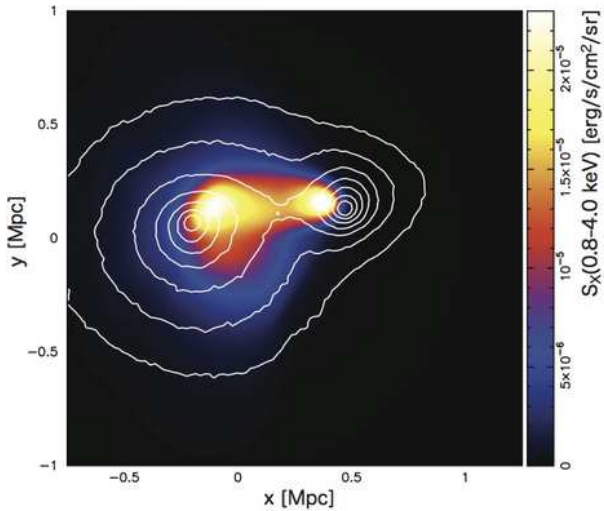}
\end{center}
\caption{X-ray surface brightness in the 0.8--4.0 keV band. The
projected mass density (dark matter + ICM) is also overlaid (white
contours). The core of the sub cluster (the right cluster) is moving
from left to right hand side and has already penetrated the main cluster
(the left cluster). \label{fig:1}}
\end{figure}

Figure~\ref{fig:1} depicts the 0.8--4.0~keV band X-ray surface
brightness map and the contours of the projected mass density (dark
matter + ICM) integrated along the line-of-sight. Our nominal run nicely
reproduces the overall structure of the 1E0657-56; for instance,
distinct two X-ray peaks, positions and shapes of shock front and
contact discontinuity, and displacement between X-ray and projected mass
density peaks associated with the sub cluster, are consistent with the
observations (\cite{mar06, clo04}).

\begin{figure}[h]
\begin{center}
\FigureFile(80mm,40mm){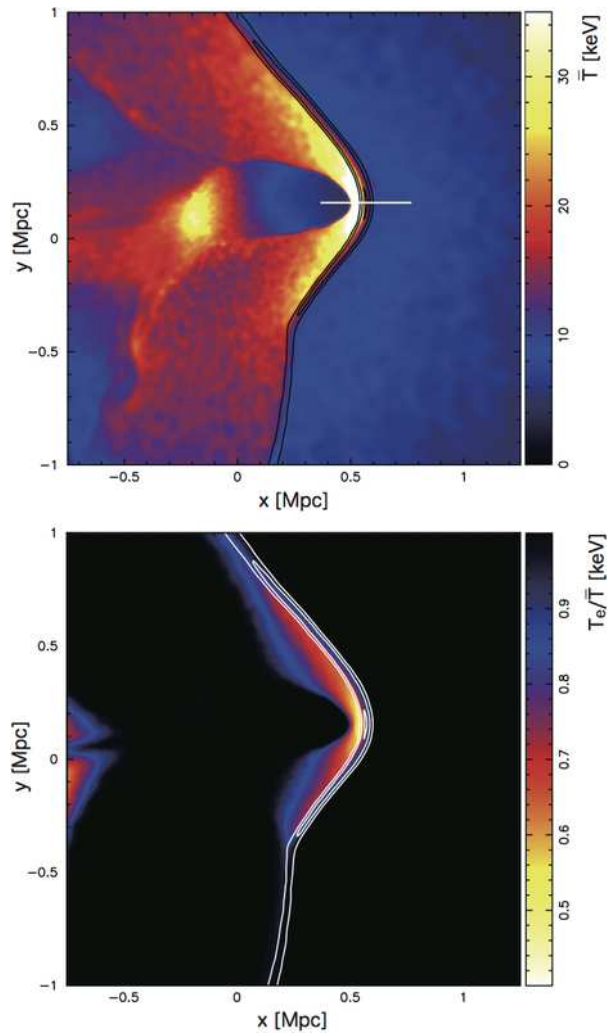}
\end{center}
\caption{The mean temperature of ICM, $\bar{T}$, (top)
and the ratio of the electron temperature relative to the mean 
temperature, $T_{\rm e}/\bar{T}$, (bottom),
on the collision plane of the two galaxy clusters.
Contours show the Mach number on the collision plane from
1.5 by 0.5, and the innermost contour represents the Mach number
of 2.5, estimated by the shock-capture method described in \citet{pfr06}.
A white horizontal line indicates the location of the 
one-dimensional profile discussed in figure~\ref{fig:7}. \label{fig:2}}
\end{figure}

Figure~\ref{fig:2} shows the temperature maps on the collision
plane of the two galaxy clusters. There exists very hot ICM heated from
$\sim 8$~keV to more than $\sim 30$~keV by a shock wave with a Mach 
number of $M\sim 2$--3 in front (the right hand side) of the cold core (the
bullet) of the sub cluster. The shock layer extends toward the outskirts of 
the clusters, which has a Mach number of $M\sim 1.5$--2.5 and heats 
the ICM to $\sim 15$--20~keV. In the two-temperature run, we can
see that the electron temperature is typically $\sim 30$--50~\% lower 
than the mean temperature behind the shock (the bottom panel of 
figure~\ref{fig:2}). 
The two-temperature structure can not be seen inside the bullet, because 
the timescale of thermal relaxation between electrons and ions are 
much shorter than the merger timescale due to the high density and low 
temperature. We have another high temperature region behind the bullet 
heated by the compressive flows.

\begin{figure}[tp]
\begin{center}
\FigureFile(80mm,40mm){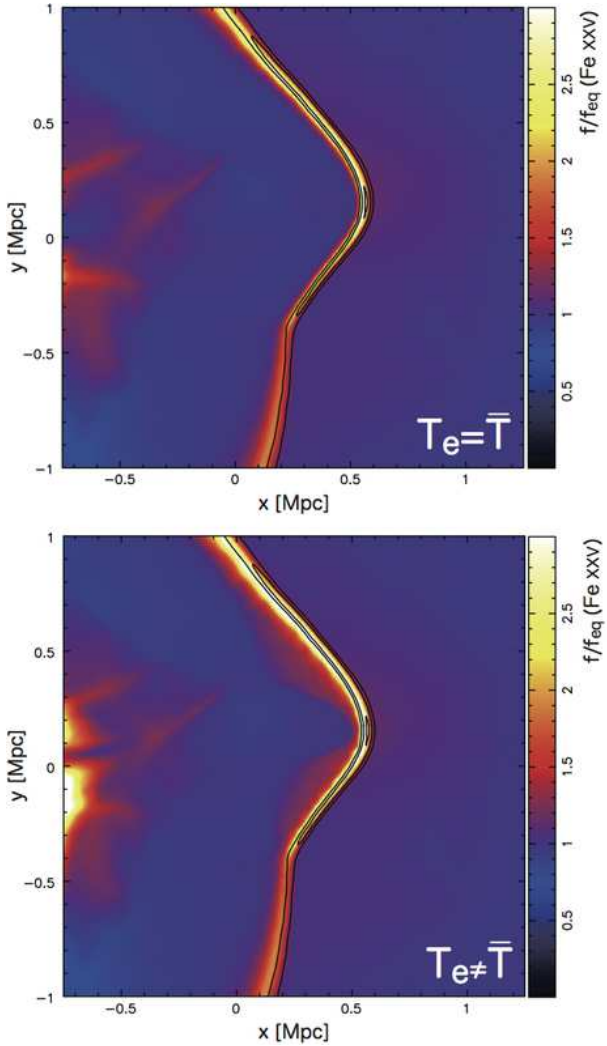}
\end{center}
\caption{The ratio of the ionization fraction of Fe\,\textsc{xxv}
 relative to that in the ionization equilibrium state
on the collision plane of the two galaxy clusters.
Top and bottom panels show the results of the single- and
two-temperature runs, respectively. Black contours are the same as 
figure~\ref{fig:2}. \label{fig:3}}
\end{figure}

\begin{figure}[tp]
\begin{center}
\FigureFile(80mm,40mm){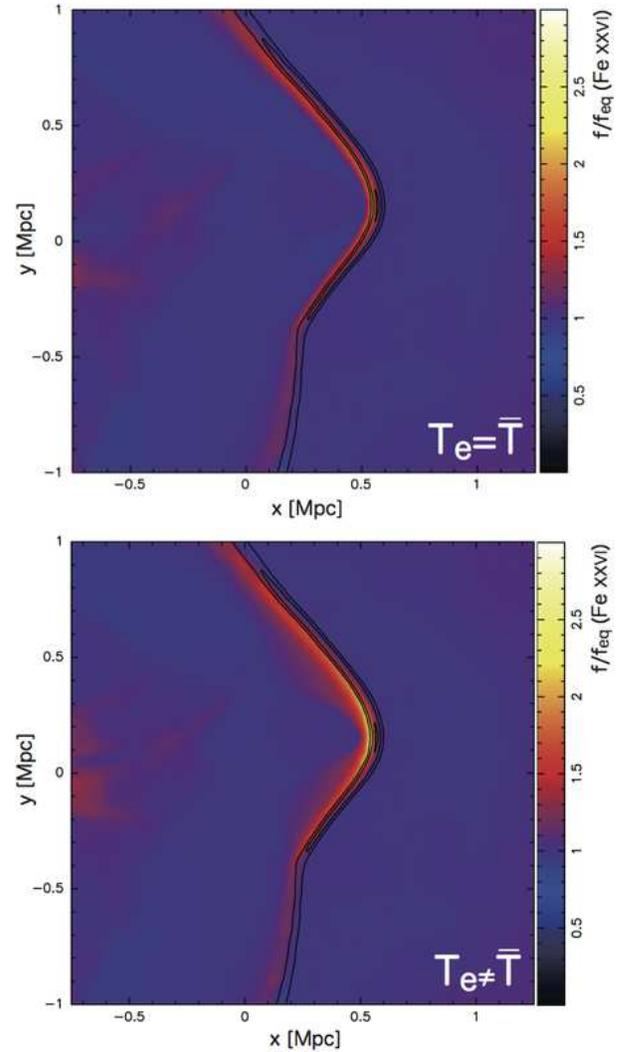}
\end{center}
\caption{Same as figure~\ref{fig:3} but for the ionization fraction of 
Fe\,\textsc{xxvi}. \label{fig:4}}
\end{figure}

Figures~\ref{fig:3} and \ref{fig:4} show the ratios of Fe\,\textsc{xxv} and
Fe\,\textsc{xxvi} fractions relative to those in the ionization
equilibrium state, respectively. Here, in the two-temperature run,
ionization fractions in the ionization equilibrium state are computed
using the mean temperature. We find
that both Fe\,\textsc{xxv} and Fe\,\textsc{xxvi} are overabundant by
1.2--3.0 behind the shock layer, and the deviations from the equilibrium
value are larger in the two-temperature run. The over-population of
Fe\,\textsc{xxv} and Fe\,\textsc{xxvi} fractions can be understood as
follows. The shock heats the ICM in the post-shock region from several
keV to a few tens keV, and both of Fe\,\textsc{xxv} and
Fe\,\textsc{xxvi} fractions in the post-shock region decrease with time
toward the equilibrium value because their fractions are maximized at
$\sim 3$~keV and $\sim 10$~keV, respectively, in the ionization
equilibrium state (see figure 4 of AY10). However, the ionization of
Fe\,\textsc{xxv} and Fe\,\textsc{xxvi} toward higher--ionized levels is
not quick enough to catch up with the ionization equilibrium
state. Therefore, these fractions are higher than that in the ionization
equilibrium state. Since the ionization rates of ions to the higher
ionization levels depend on the electron temperature, and the rates are
significantly smaller in the two-temperature run, a delay of the heating
of electrons at the shock leaves these fractions higher than that in the
single-temperature run.

\begin{figure}[tp]
\begin{center}
\FigureFile(80mm,40mm){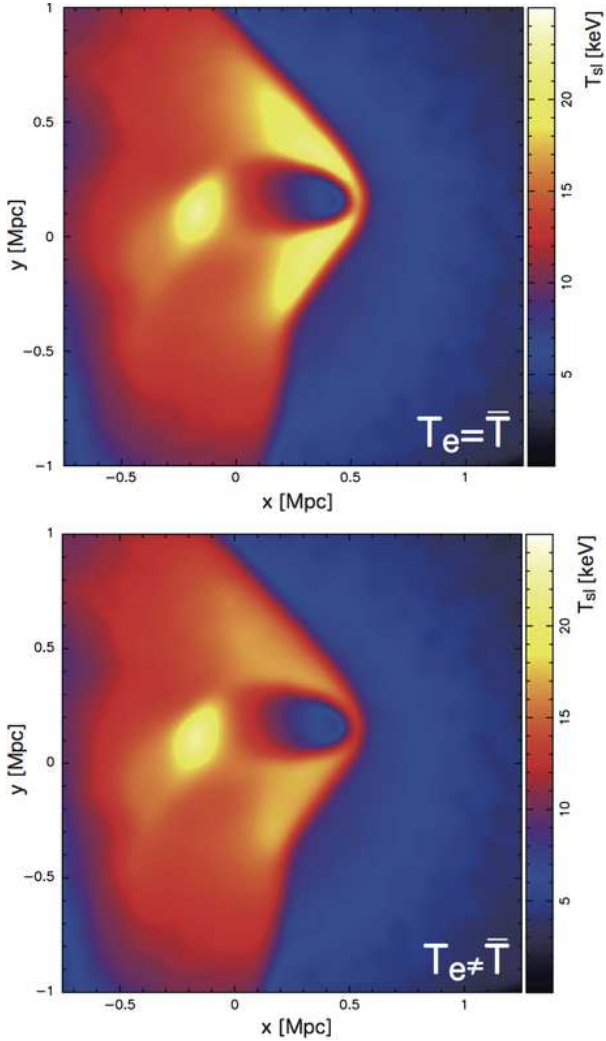}
\end{center}
\caption{The spectroscopic-like temperature (\cite{maz04}).
Top and bottom panels show the results of the single- and
two-temperature runs, respectively. \label{fig:5}}
\end{figure}

\begin{figure}[tp]
\begin{center}
\FigureFile(80mm,40mm){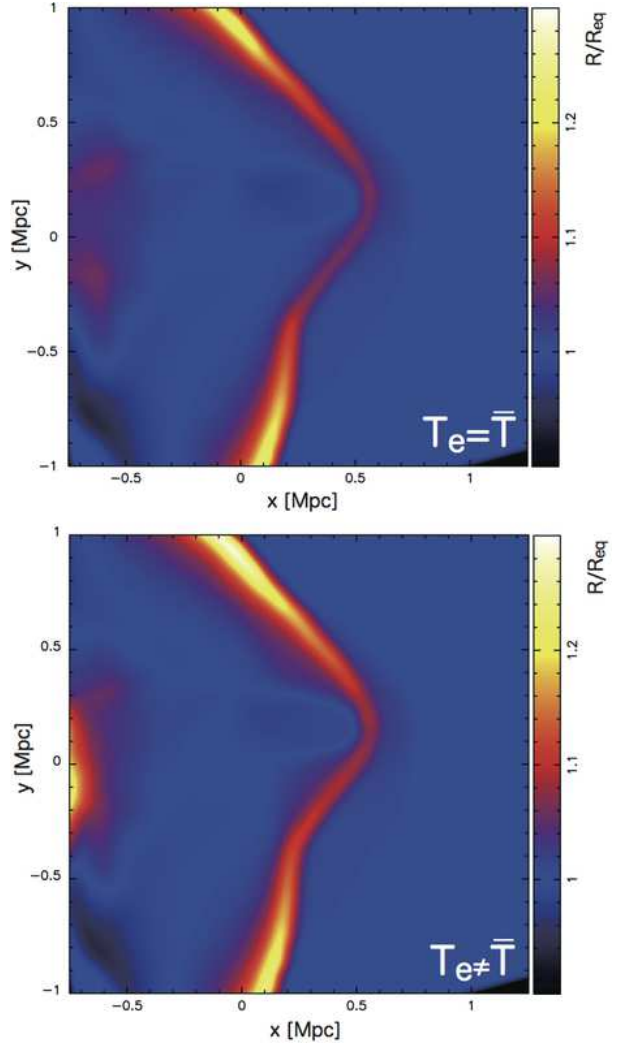}
\end{center}
\caption{$R/R_{\rm eq}$ (see equation~(\ref{eq:RReq})).
Top and bottom panels show the results of the single- and
two-temperature runs, respectively. \label{fig:6}}
\end{figure}

To see an observational appearance of the result of the nominal runs, we
calculate the spectroscopic-like temperature ($T_{\rm sl}$,
\cite{maz04}) and the ratio, $R/R_{\rm eq}$, as an observational tracer
for the non-equilibrium ionization state (AY10), where $R$ is the X-ray
intensity ratio between rest-frame 6.6--6.7~keV and 6.9--7.0~keV energy
bands in which the Fe\,\textsc{xxv} K$\alpha$ and Fe\,\textsc{xxvi}
K$\alpha$ emissions are predominant, respectively;
\begin{equation}
\label{eq:RReq}
R=\frac{I(6.6-6.7~{\rm keV})}{I(6.9-7.0~{\rm keV})},
\end{equation}
and $R_{\rm eq}$ is the intensity ratio defined above but in the
ionization equilibrium state. It should be noted that the quantity,
$R/R_{\rm eq}$, is independent on the ICM metallicity. 
The results are shown in figures \ref{fig:5} and \ref{fig:6}. In terms of the
spectroscopic-like temperature, the temperature of the ICM at the shock in
front of the bullet is roughly $\sim 15$~keV in the two-temperature run,
and lower than in the single-temperature run. $R/R_{\rm eq}$ departs
from unity ($\sim 1.1$--1.2) both in the single- and two temperature runs, 
though its difference between the two runs is not significant. In both single- and
two-temperature runs, the deviation of the intensity ratio is more
significant in the outskirts than in the central part of the clusters,
since the timescale to reach the ionization equilibrium state is longer
in the lower-density regions.

\subsection{Detailed Structures of Bullet and Shock}

The shock front ahead of the bullet is suited to the study of the
non-equilibrium ionization state and the two-temperature structure of
the ICM, since the strong shock with a Mach number of $3.0\pm0.4$ is
reported \citep{mar06} and the observational signatures of the departure
from the ionization equilibrium and the two-temperature structure are
expected to be relatively strong. In this subsection, we present the
detailed analyses of the shock ahead of the bullet, and how such
observational signatures can be detected in X-ray observations.

\begin{figure}[tp]
\begin{center}
\FigureFile(80mm,40mm){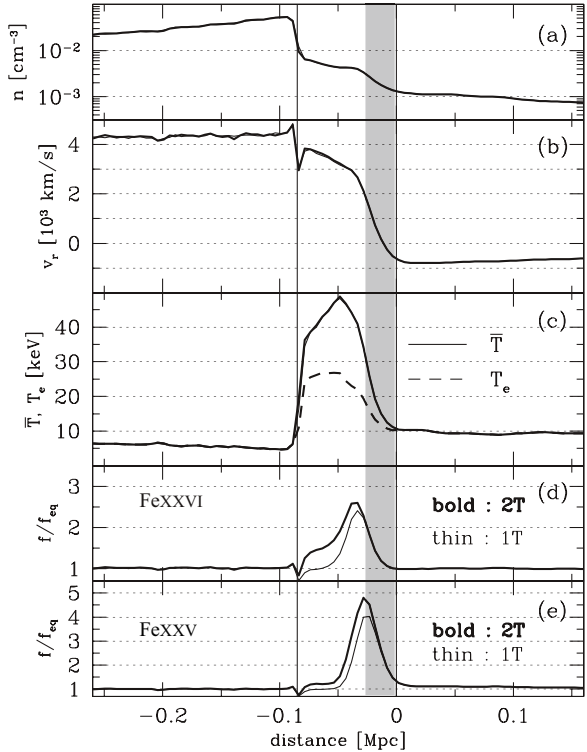}
\end{center}
\caption{One dimensional profiles across the shock front on the
collision plane. Thin and thick lines show the results of the single-
and two-temperature runs, respectively. Panels from the top to the
bottom show (a) the ICM density, (b) the ICM velocity along the axis,
(c) the average temperature (solid line) and the electron temperatures
in the two-temperature run (dashed line), (d) the ratio of ionization
fractions of Fe\,\textsc{xxv} relative to that in the ionization
equilibrium state, and (e) the ratio for Fe\,\textsc{xxvi}. Two vertical
lines indicate the location of the shock front (right) and the contact
discontinuity (left). The coordinate of the horizonal axis shows the
distance measured from the shock front.
\label{fig:7} }
\end{figure}

Figure~\ref{fig:7} shows the one dimensional profiles of the ICM across
the bullet and the shock front on a line on the collision plane, the
location of which is indicated by a white line in
figure~\ref{fig:2}. We can see that the shock velocity, the relative
velocity of the ICM between pre- and post-shock regions, is about
$5000$~km/s (figure~\ref{fig:7}(b)), which is consistent with the
previous simulations (\cite{tak06, mil07, spr07, mas08}), The Mach
number of the shock is $\sim 2.8$, and consistent with the estimate
based on X-ray observations by {\it Chandra} \citep{mar06}.

In the two-temperature run, we find that the electron temperature
between the shock front and the contact discontinuity is 
$T_{\rm e} \sim 25$~keV and much lower than the mean 
temperature, $\bar{T} \sim 45$~keV.  In the single-temperature run, 
the temperature of the ICM is almost the same as the mean temperature in 
the two-temperature run, and the profiles of the ICM density and velocity 
are also unchanged regardless of the thermal relaxation model adopted. 
This means that the difference in the cooling rate due to the difference 
in the electron temperature does not affect the overall hydrodynamic 
properties of the ICM. 

We can see that the ionization state of iron strongly deviates from
the ionization equilibrium state between the shock front and the contact
discontinuity, and the ratios of Fe\,\textsc{xxv} and Fe\,\textsc{xxvi}
fractions relative to those in the ionization equilibrium state reach
$\sim 5$ and $\sim 2.5$, respectively (figures \ref{fig:7}(d) and
\ref{fig:7}(e)). The amount of the departure from the
ionization equilibrium state depends on the adopted thermal relaxation
model and larger in the two-temperature run.

\begin{figure}[tp]
\begin{center}
\FigureFile(80mm,40mm){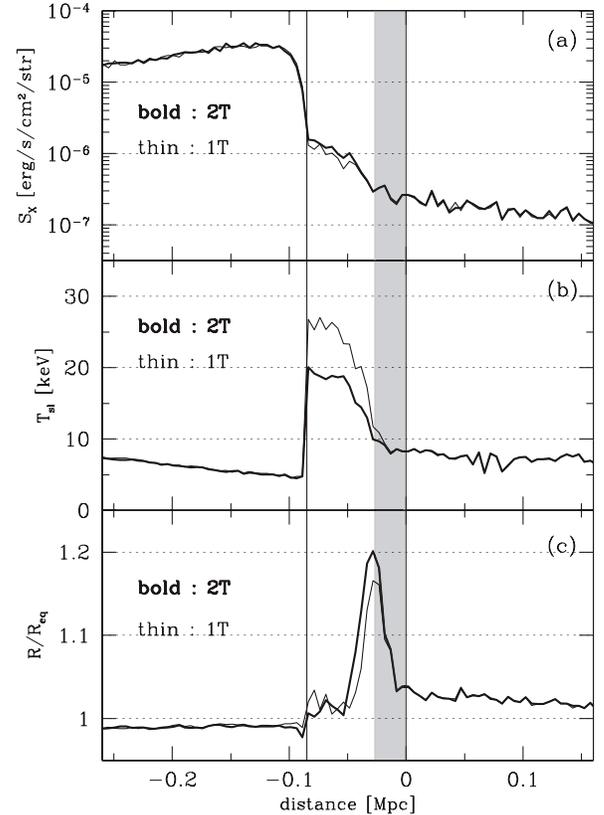}
\end{center}
\caption{Same as figure~\ref{fig:7} except that panels from the top
to the bottom show (a) the surface brightness in the 0.8--4.0~keV energy
band, (b) the spectroscopic-like temperature, and (c) $R/R_{\rm eq}$
projected along the line-of-sight, respectively. \label{fig:8}}
\end{figure}

Figure~\ref{fig:8} shows the profiles of X-ray surface brightness in the
0.8--4.0~keV band (panel (a)), spectroscopic-like temperature (panel
(b)), and $R/R_{\rm eq}$ (panel (c)) projected along the line-of-sight 
at the same
region as figure~\ref{fig:7}. In these projected quantities, the
signatures of the two-temperature structure and the non-equilibrium
ionization state are contaminated by the contributions of the foreground
and background ICM. Nevertheless, we can see such signatures in
figure~\ref{fig:8} to some extent. In the two-temperature run, the
maximum spectroscopic-like temperature between the shock front and the
contact discontinuity is $\sim 20$~keV and significantly lower than that
in the single-temperature run ($\sim 25$~keV). $R/R_{\rm eq}$ also
deviates from unity at the same region, indicating the non-equilibrium
ionization state in this region, though the difference between the
single- and two-temperature runs is not significant. 

It should be noted that, due to the poor capability of the SPH scheme to
resolve the shock front, the discontinuities of physical quantities such
as density and temperature across the shock front are numerically
smeared in our simulations, and that the actual jumps of density,
velocity and temperature of the ICM should be sharper in
``reality''. The shaded regions in figures \ref{fig:7} and \ref{fig:8}
indicate the locations where such numerical smearing is effective.  If
the jumps of physical quantities at the shock are ideally resolved, it
is expected that the difference in temperature between electrons and
ions at the post-shock regions and the departure from the ionization
equilibrium state are larger at the post-shock region. Therefore, the
difference of projected profiles of the spectroscopic-like temperature
between the single- and two-temperature runs would be larger, and thus
the difference of $R/R_{\rm eq}$ between the single- and two-temperature
runs would be also more significant. 

\section{Dependence on the Initial Condition}

In this section, in order to figure out parameter dependence of our
results, we compare various merger simulations with different sets of
parameters.

First, we briefly summarize morphological differences of the simulated 
clusters. The followings are notable dependences on the initial parameters
that we confirmed:

\begin{itemize}
 \item An overall structure of the observed features such as the bullet, the
       shock layer, and the peaks of X-ray surface brightness and
       projected mass density are best reproduced with the concentration
       parameter, $c=5$, of the main cluster. The runs with $c=3$ (e.g.,
       \cite{spr07}) give similar results, but produce X-ray and
       projected mass density maps of the main cluster slightly more
       elongated along north-south direction. Difference in the simulations
       with $c=5$ and those with $c=6$ adopted by \citet{mas08} is
       almost negligible.

 \item The displacement between the peaks of X-ray surface brightness
       and projected mass density in the sub cluster (the bullet) are
       best reproduced with the concentration parameter, $c=8$, of the
       sub cluster (\cite{mas08}).  For $c>8$, the bullet gets close to
       the projected density peak, and the displacement becomes smaller.

 \item As pointed out by \citet{mas08}, a non-zero impact parameter is
       necessary to reproduce the asymmetric X-ray surface
       brightness. However, the runs in which the sub-cluster gets
       through the south side of the main cluster can not simultaneously
       reproduce the observed X-ray surface brightness and projected
       mass density. Therefore, we adopt the sub-cluster's orbit which
       passes through the north side of the main cluster.
 \item The initial relative velocity of 3000 km/s is required to
       reproduce the shock layer with a Mach number of $M\sim 3$ in
       front of the bullet.  Similar requirement has been also indicated
       in other numerical studies on the bullet cluster
       (\cite{tak06,mil07,spr07,mas08}). According to \citet{hay06}, an
       encounter with the virial velocity of the main cluster, $V_{\rm
       200}\sim 1700$~km/s, and the relative velocity, $V_{\rm sub}\sim
       3000$~km/s, is probable with a relative likelihood of $f\sim
       0.03$, based on the cosmological structure formation simulations
       (see equation (1) of Hayashi \& White 2006, where we adopted
       $v_{\rm 10percent}=1.55$, and $\alpha=3.3$).  However, it should
       be noted that the probability for such a high initial relative
       velocity is still controversial: \citet{lk10} recently claimed
       that the initial relative velocity of 3000~km/s at $R_{200}$ is
       very unusual with the prediction of the concordance $\Lambda$CDM
       model. Even if we admit a $M\sim 2.5$ shock as the lower limit of
       the observed Mach number, the required initial relative velocity
       is $\sim$2500~km/s, and is still peculiarly high.
 \item The contrast of the peak brightness relative to that of the
       ambient gas is best reproduced with the initial mass ratio of
       $6:1$ between the main and the sub clusters. With decreasing the
       mass ratio, the X-ray peak of the main cluster becomes dimmer
       (see e.g., $3:1$ runs of \cite{mas08}). Note that the mass ratio
       also affects the Mach number of the shock; the run with the mass
       ratio of $4:1$ produces the shock with the Mach number larger
       than 3 for the initial relative velocity of 3000 km/s, and allows
       us to adopt smaller initial relative velocity to reproduce the
       $M\sim 3$ shock.
 \end{itemize}

Using the various merger simulations with different sets of parameters,
let us evaluate a possible range of the non-equilibrium effects in the
1E0657-56 system. Based on the simulations in this and previous works
(\cite{spr07}, \cite{mas08}), we consider the results with the initial
mass ratio, $4:1$, $6:1$, or $10:1$, and the concentration parameter
$c=$ 3 or 5 for the main cluster and 8 or 10 for the sub cluster, which
are consistent with the observational features of the 1E0657-56 system
to some extent. To identify the snapshot that corresponds to the
``current'' epoch in the time sequences of the simulations, we adopt the
separation of two peaks of the projected mass density. The other
parameters such as the mass of the main cluster and the impact parameter
are the same as our nominal run except that we use one eighth of the
number of particles (the smoothing length is thus twice as large). We
adopt the same initial relative velocity of 3000~km/s regardless of
the mass ratio, since, in our prescription to determine the initial
relative velocity described in \citet{sar02}, it is proportional to the
square root of the mass ratio and has weak dependence on the mass ratio.

In table~1, we present the physical properties of the shock front
associated with the bullet for the runs with the simulation parameters
different from our nominal run. We find that $R/R_{\rm eq}=1.1$--1.4 in
the two-temperature runs, with the Mach number ranging from $M=2.5$ to
3.4. We also find that $\bar{T}=33$--50~keV and $T_{\rm e}=9$--24~keV,
or $T_{\rm e}/\bar{T}=0.2$--0.7.  In other words, the observed electron
temperature at the shock lower than $\sim 25$~keV suggests that the
Coulomb collision is a predominant process of thermal relaxation between
electrons and ions and that there exists the two-temperature structure
at the shock front. On the other hand, the observed electron temperature
higher than $\sim 30$~keV prefer the instantaneous thermal relaxation at
the shock front.

The {\it Chandra} observation suggests that the deprojected electron
temperature behind the shock is over $\sim 30$~keV (\cite{mar06}), which
is higher than the constraint for the existence of the two-temperature
structure we obtained in the two-temperature runs. This suggests that,
although there are still uncertainties due to the spatial resolution in
both observations and simulations, the instantaneous thermal relaxation
model is preferable for reproducing the observed temperature structure
at the shock. \citet{mar06} discussed that the simple model of
electron-ion thermal relaxation only through the Coulomb scattering is
inconsistent with the observed data and is excluded at a 95\% confidence
level, implying that the thermal equilibration between electrons and
ions should be much faster than that by the Coulomb scattering. We,
however, should notice that the accuracy of the observed electron
temperature by {\it Chandra} is somewhat limited in the high energy band
more than $\sim 10$~keV. Thus, this issue should be addressed by the
observations with future X-ray satellites such as {\it Astro-H} with
better sensitivity in high energy band. Another plausible way to measure
the temperature of very hot ICM should be a joint analysis with the X-ray
and SZ data \citep{kit04}. The analysis of the Bullet cluster 
that is based on the high-resolution X-ray and SZ observations such as
{\it Chandra} and {\it Herschel-SPIRE} will provide us new constraints on
very hot thermal and/or non-thermal particle populations
(Prokhorov, Akahori, Yoshikawa, et al., in preparation).

\begin{table}
\caption{Physical properties of the shock in front of the bullet for the
runs ``c$X$c$Y$m$Z$'' where $X$ and $Y$ denotes the concentration
parameters of the main and sub clusters, respectively, and the mass
ratio of the two clusters is $Z:1$. $\bar{T}$ and $T_{\rm e}$ are in
units of keV.\label{tab:1}}
\begin{center}
\begin{tabular}{lcccccccccc}
 \hline
 \noalign{\smallskip}
 run ID & $M$ & $\bar{T}$ & $T_{\rm e}$ & ${T_{\rm e}/\bar{T}}$ & ${R/R_{\rm eq}}$ \\
 \noalign{\smallskip}
 \hline
 \noalign{\smallskip}
c03c08m04 & 3.0 & 44 & 21 & 0.47 & 1.25\\
 \noalign{\smallskip}
c03c08m06 & 3.0 & 43 & 21 & 0.50 & 1.26 \\
 \noalign{\smallskip}
c03c08m10 & 2.6 & 33 & 20 & 0.62 & 1.24 \\
 \noalign{\smallskip}
c03c10m04 & 3.1 & 49 & 20 & 0.40 & 1.29 \\
 \noalign{\smallskip}
c03c10m06 & 3.4 & 41 & 13 & 0.31 & 1.36 \\
 \noalign{\smallskip}
c03c10m10 & 3.2 & 34 & 8.3 & 0.26 & 1.19 \\
 \noalign{\smallskip}
c05c08m04 & 2.9 & 47 & 24 & 0.51 & 1.27 \\
 \noalign{\smallskip}
c05c08m06 & 3.0 & 43 & 23 & 0.54 & 1.19 \\
 \noalign{\smallskip}
c05c08m10 & 2.5 & 34 & 22 & 0.65 & 1.18 \\
 \noalign{\smallskip}
c05c10m04 & 3.1 & 50 & 22 & 0.43 & 1.23 \\
 \noalign{\smallskip}
c05c10m06 & 3.0 & 49 & 16 & 0.34 & 1.14 \\
 \noalign{\smallskip}
c05c10m10 & 3.2 & 42 & 9.9 & 0.23 & 1.16 \\
 \noalign{\smallskip}
 \hline
\end{tabular}
\end{center}
\end{table}

\section{Effect of Radiative Cooling}

Finally, we discuss the effect of radiative cooling on the properties of the 
simulated bullet clusters.

As partly discussed in section 3.2, the thermal relaxation model does not 
significantly change the overall hydrodynamic properties of the ICM. 
This is because the radiative cooling is inefficient in the pre-shock region 
and the shock layer, where radiative cooling timescale is longer than the 
age of the universe, regardless of the difference in the cooling rate due 
to the difference in the electron temperature. As for the bullet, radiative 
cooling is efficient, but there are no two-temperature structure due to a 
short relaxation timescale between electrons and ions, so that the effect 
of radiative cooling is identical there. Therefore, morphology of the X-ray 
surface brightness does not depend significantly on the thermal relaxation
model, and thus we basically obtain the same best fit parameters for 
different thermal relaxation model.

\begin{figure}[tp]
\begin{center}
\FigureFile(80mm,40mm){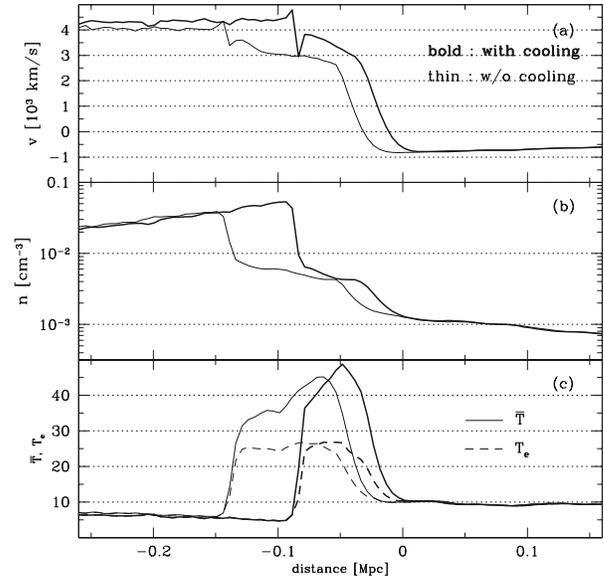}
\end{center}
\caption{One dimensional profiles across the bullet and the shock on the
collision plane for the two-temperature runs. Thin and thick lines show
the results of the non-cooling and cooling runs, respectively. Panels
from the top to the bottom shows (a) the ICM velocity along the axis,
(b) the ICM density, and (c) the average temperature (solid) and the
electron temperature (dashed).\label{fig:9} }
\end{figure}

Considering the effect of radiative cooling is, however, quite important 
to determine the location of the bullet in the 1E0657-56 system. 
In order to demonstrate the effect, we carried out the control run in 
which we neglect the effect of radiative cooling, and compared it with 
the cooling run that we have shown. The results after 1.12 Gyr has 
elapsed for the two-temperature runs are shown in figure \ref{fig:9}, 
where the axes of the profile for two runs are the same as that shown 
in figure \ref{fig:2}. We can see a delay of the propagation of the 
bullet and the shock in the non-cooling run. Moreover, the width 
between the shock front and the contact discontinuity becomes wider 
in the non-cooling run than that in the cooling run. This may be ascribed 
to the fact that the gravitational binding force is weaker, i.e. ram 
pressure becomes more significant, in the non-cooling run, since the 
bullet less concentrates compared with the cooling run (by a factor of 
$\sim$ 1.6 at the present). Our result suggests that considering the 
effect of radiative cooling is important to correctly constrain the initial 
relative velocity from the measurement of the width between the shock 
front and the contact discontinuity \citep{mas08}.

\section{Summary and Conclusion}

We conduct a set of $N$-body/SPH simulations of merging galaxy clusters
which reproduce observed X-ray surface brightness and projected mass density
distribution of the 1E0657-56 system to investigate the non-equilibrium
ionization state and electron-ion two-temperature structure in the 
1E0657-56 system, and to assess their detectability in future X-ray missions.
Different from previous numerical works of the 1E0657-56 system,
we, for the first time, relax the both assumptions of ionization equilibrium
and thermal equipartition between electrons and ions.

Our nominal run, which reproduces the various observed features of the
1E0657-56 system fairly well, is an merger of two galaxy clusters with a
mass ratio of 6:1 and an initial relative velocity of 3000~km/s. It is
found that departure from the ionization equilibrium is significant at
the shock front with a Mach number of $M\sim 3$ associated with the
penetrating core of the sub cluster. When considering only the Coulomb
scattering as a physical process for the thermal relaxation between
electrons and ions, we find that the two-temperature structure, or the
difference in temperature between electrons and ions, are also
significant at the shock front.

Comparisons of the X-ray and column density profiles with those 
in the observations gave possible range of simulation parameters. 
Within the range, at the shock associated with the bullet we found 
that $R/R_{\rm eq}$ (equation (\ref{eq:RReq})) clearly deviates 
from unity and the electron temperature is much lower than the mean 
temperature of the ICM in the two-temperature run. Effect of radiative 
cooling is important to determine the location of the bullet, where 
the overall hydrodynamic properties of the ICM is not significantly 
altered by the different efficiency of radiative cooling due to the 
different temperature structure.

It is found that, in the two-temperature run, the spectroscopic-like
temperature projected along the line-of-sight between the shock front
and the contact discontinuity is significantly lower than that in the
single-temperature run, which suggests that precise measurements of the
maximum electron temperature in this region could provide strong
constraints on thermal relaxation processes between electrons and ions.
Such observations can be available with a high sensitivity X-ray
detector in high energy band and also with a good spatial resolution to
resolve the relatively tiny area between the shock front and the contact
discontinuity. Observations of thermal Sunyaev--Zel'dovich effect with
relatively good spatial resolution can provide an independent
measurement of electron temperature in this region.
A good spectroscopic resolution in X-ray observations is also important
in studying the two-temperature structure of the ICM, because the ion
temperature is only estimated from the detailed line profiles of
emission lines, and would be achieved by X-ray calorimeters on board the
forthcoming satellites. Our results also suggest that the current X-ray
observations can lead to a biased result on the estimation of the
metallicity if one assumes that the ICM is in the ionization equilibrium
state, because the separate detection of emission lines of
Fe~\textsc{xxv} and Fe~\textsc{xxvi} is still difficult with the current
X-ray observational facilities due to a lack of the spectroscopic
resolution of the CCDs. Therefore, a good spectroscopic resolution is
also needed to correctly measure the metallicity in the shock heated
ICM.

\bigskip

This work was supported in part by Grant-in-Aid for Scientific Research
(S) (20224002), for Scientific Research (A) (20340041), for Young
Scientists (Start-up) (19840008) and for Challenging Exploratory
Research (21654026) from JSPS.  Numerical simulations for the present
work had been carried out under the ``Interdisciplinary Computational
Science Program'' in Center for Computational Sciences, University of
Tsukuba.  TA was supported in part by Korea Science and Engineering
Foundation (R01-2007-000-20196-0) and by the National Research 
Foundation of Korea (2007-0093860).

\end{document}